\documentclass[aip,jcp,amsmath,amssymb,reprint]{revtex4-1}

\usepackage{amsmath}
\usepackage{amsfonts}
\usepackage{amsthm}

\usepackage{natbib}
\usepackage{textcomp}
\usepackage[letterpaper,margin=3cm]{geometry}
\usepackage{graphicx}
\usepackage{float}
\usepackage{hyperref}

\allowdisplaybreaks

\begin{document}
\title{Stochastic Density Functional Theory: Real- and Energy-Space Fragmentation for Noise Reduction}

\author{Ming Chen}
\affiliation{Department of Chemistry, University of California, Berkeley, California 94720, USA}
\affiliation{Materials Sciences Division, Lawrence Berkeley National Laboratory, Berkeley, California 94720, USA}

\author{Roi Baer}
\affiliation{Fritz Haber Center of Molecular Dynamics and Institute of Chemistry, The Hebrew University of Jerusalem, Jerusalem, 91904 Israel}

\author{Daniel Neuhauser}
\affiliation{Department of Chemistry and Biochemistry, University of California, Los Angeles, California 90095, USA}

\author{Eran Rabani}
\affiliation{Department of Chemistry, University of California, Berkeley, California 94720, USA}
\affiliation{Materials Sciences Division, Lawrence Berkeley National Laboratory, Berkeley, California 94720, USA}
\affiliation{The Raymond and Beverly Sackler Center of Computational Molecular and Materials Science, Tel Aviv University, Tel Aviv 69978, Israel}

\begin{abstract}
Stochastic density functional theory (sDFT) is becoming a valuable tool for studying ground state properties of extended materials. The computational complexity of describing the Kohn-Sham orbitals is replaced by introducing a set of random (stochastic) orbitals leading to linear and often sub-linear scaling of certain ground-state observable at the account of introducing a statistical error. Schemes to reduce the noise are essential, for example, for determining the structure using the forces obtained from sDFT. Recently we have introduced two embedding schemes to mitigate the statistical fluctuations in the electron density and resultant forces on the nuclei. Both techniques were based on fragmenting the system either in real-space or slicing the occupied space into energy windows, allowing for a significant reduction of the statistical fluctuations.  For chemical accuracy further reduction of the noise is required, which could be achieved by increasing the number of stochastic orbitals. However, the convergence is relatively slow as the statistical error scales as $1/\sqrt{N_\chi}$ according to the central limit theorem, where $N_\chi$ is the number of random orbitals. In this paper we combined the aforementioned embedding schemes and introduced a new approach that builds on overlapped fragments and energy windows. The new approach significantly lowers the noise for ground state properties such as the electron density, total energy, and forces on the nuclei, as demonstrated for a G-center in bulk silicon. 
\end{abstract}
\maketitle

\section{Introduction}
Kohn-Sham (KS) density functional theory~\cite{PhysRev.136.B864,PhysRev.140.A1133} (DFT) is 
widely used to study a wide range of systems due to its capability of quantitatively predicting ground state properties at a moderate computational cost of $O(N_e^3)$, where $N_e$ is the number of electrons. While this moderate scaling allows for an efficient description of the ground state of molecules and bulk structures with periodic boundary conditions, the application to systems containing $10^4$ electrons or more, such as nanostructures,\cite{Aliano2012} complex materials,\cite{R0953-8984-14-11-313} and large biomolecules,\cite{Cole_2016} is still a severe challenge for today's DFT implementations. Linear-scaling methods for DFT based on dividing the entire system into subsystems~\cite{PhysRevLett.66.1438,PhysRevB.44.8454,PhysRevB.53.12713} require proper treatment of the boundaries between the fragments,\cite{PhysRevB.53.12713,doi:10.1063/1.470549,doi:10.1021/ct9001784,doi:10.1021/cr500502v,doi:10.1063/1.3582913,doi:10.1063/1.3659293} while methods that rely on the sparsity of density matrix (DM)~\cite{PhysRevB.47.9973,PhysRevB.48.14646,GOEDECKER1995261,PhysRevB.51.10157,PhysRevLett.76.3168,PhysRevLett.79.3962,PhysRevB.58.12704} suffer from slow convergence for systems with small fundamental band gaps.\cite{PhysRevLett.79.3962} 

We have recently introduced an alternative linear-scaling approach to DFT which, does not rely on the  partitioning of the system into subsystems, nor does it depend on the sparsity of the density matrix.\cite{PhysRevLett.111.106402}
Instead, it utilizes stochastic orbitals, which are random linear combinations of deterministic KS orbitals to calculate the electron density and other ground-state properties. In practice, the required number of stochastic orbitals does not increase with the system size for evaluating many ground state properties,
\cite{PhysRevLett.111.106402,doi:10.1063/1.5064472} leading to linear scaling or even sub-linear scaling DFT.\cite{PhysRevLett.111.106402,doi:10.1063/1.5064472} In stochastic DFT, linear scaling is achieved by introducing a statistical error in the density and related observables, which according to the central limit theorem, decreases rather slowly with the number of stochastic orbitals, $N_{\chi}$, limiting the efficiency and accuracy of the method. Therefore, developing noise reduction schemes for sDFT is essential for achieving chemical accuracy without the need to dramatically increase $N_{\chi}$.

One approach for reducing the noise in sDFT is based on dividing the entire system into fragments.  
The entire system's density is then given as a sum of the fragment densities and a correction term sampled using stochastic orbitals. When the sum of the fragment densities provides a good approximation of the total system's density, the correction term is small, leading to significant reductions of the noise for the electron density, energy, and forces on the nuclei. This approach has been illustrated for systems with open boundary conditions~\cite{doi:10.1063/1.4890651,doi:10.1063/1.4984931,doi:10.1002/wcms.1412} as well as for periodic boundary conditions.\cite{doi:10.1063/1.5064472} For the latter case, we used overlapped fragments to ensure a reasonable estimate of both the density and the density matrix (this approach was referred to as ``overlapped embedded fragmented stochastic density functional theory'' (o-efsDFT).\cite{doi:10.1063/1.5064472}) Recently, we introduced an alternative technique to mitigate the statistical noise, referred to as ``energy-window sDFT'' (ew-sDFT),\cite{doi:10.1063/1.5114984} where the occupied space was divided into energy-resolved subspaces (``energy windows'') and the contribution to the density for each window can be calculated simultaneously. This method reduces the statistical noise in the density and the nuclei forces, but not in the total electronic energy.\cite{doi:10.1063/1.5114984}

In this paper, we combine the overlapped embedded fragmented scheme with the energy window scheme. Noise reduction is obtained by projecting both the system density matrix and the fragment density matrix onto fixed energy windows. The proposed energy window embedded fragmented stochastic DFT (ew-efsDFT) approach reduces the noise in the electron density, total energy, and forces on the nuclei and the total computational time by more than an order of magnitude compared to ew-sDFT or o-efsDFT, as illustrated for a G-center embedded in bulk silicon. The noise reduction is crucial for obtaining structural information with chemical accuracy using only several tens of stochastic orbitals, as will be shown in a proceeding publication.\cite{structmin} 

The manuscript is organized as follows: In Sec~\ref{sec:sdft}, we briefly review the sDFT. In Sec.~\ref{sec:nrs}, we present the o-efsDFT and ew-sDFT methods, both central to the development of the current noise reduction scheme. In Sec.~\ref{sec:ew-efsdft}, we provide the details of the proposed ew-efsDFT and a summary of the algorithm. Assessment of the new approach for a challenging G-center embedded in bulk silicon is presented in Sec.~\ref{sec:results} alongside a discussion of the computational complexity and cost of the ew-efsDFT.  Finally, in Sec.~\ref{sec:conclusions}, we summarize the main developments.

\section{Stochastic Density Functional Theory}
\label{sec:sdft}
Consider an extended system described by KS-DFT, with a KS Hamiltonian ($\hat{h}_{\mathrm{KS}}$) given by:
\begin{equation}
\hat{h}_{\mathrm{KS}}=\hat{t}+\hat{v}_{\text{nl}}+\hat{v}_{\text{loc}}+\hat{v}_{\text{H}}[\rho]+
\hat{v}_{\text{xc}}[\rho],
\end{equation}
where $\hat{t}$, $\hat{v}_{\text{nl}}$, $\hat{v}_{\text{loc}}$, $\hat{v}_{\text{H}}[\rho]$, and $\hat{v}_{\text{xc}}[\rho]$ are the operators of the kinetic energy, the non-local pseudopotential energy, the local pseudopotential energy, the Hartree energy, and the exchange-correlation energy, respectively. The Hartree and exchange-correlation terms depend on the electron density, $\rho(\mathbf{r})$, which is formally given by (we assume closed-shell and ignore spin-orbit couplings for simplicity):
\begin{align}
\rho(\mathbf{r}) & =2\mathrm{Tr}\left(\hat{\rho}\delta(\mathbf{r}-\hat{\mathbf{r}})\right) \nonumber \\
& = \lim_{\beta\rightarrow\infty}2\mathrm{Tr}\left(\theta_{\beta}(\hat{h}_{\text{KS}},\mu)
\delta(\mathbf{r}-\hat{\mathbf{r}})\right),
\label{eq:rho-det}
\end{align}
where $\hat{\rho}=\theta_{\beta}(\hat{h}_{\text{KS}},\mu)$ is the one-body density matrix, $\theta_{\beta}(x,\mu)=1/(1+e^{\beta(x-\mu)})$ is the Fermi-Dirac distribution function parametrized by the inverse temperature ($\beta$) and the chemical potential ($\mu$) tuned to give the number of electrons, $N_{\text e}=\int d{\bf r} \rho(\mathbf{r})$.  Other smooth functions to approximate a step function can also be used instead of $\theta_{\beta}(x,\mu)$. In KS-DFT, the electron density can also be written in terms of the KS orbitals (eigenstates of the KS Hamiltonian), $\phi_i(\mathbf{r})$: 
\begin{equation}
\rho(\mathbf{r})=2\sum_i^{N_{\text{occ}}}|\phi_i(\mathbf{r})|^2,
\end{equation}
where $N_{\text{occ}}$ is the number of occupied orbitals. 

In sDFT, the trace in Eq.~(\ref{eq:rho-det}) is replaced by averaging the expectation value of $\theta_{\beta}(\hat{h}_{\text{KS}},\mu) \delta(\mathbf{r}-\hat{\mathbf{r}})$:
\begin{equation}
\rho(\mathbf{r})=2\left\langle\left\langle\chi\middle|\theta_{\beta}(\hat{h}_{\text{KS}},\mu)
\delta(\mathbf{r}-\hat{\mathbf{r}})\middle|\chi\right\rangle\right\rangle_\chi,
\label{eq:rho-sto}
\end{equation}
where $|\chi\rangle$ is a stochastic  orbital and $\langle\cdots\rangle_\chi$ implies averaging over an ensemble of stochastic orbitals. The stochastic orbitals are represented on a real-space grid with $N_{\text g}$ grid points; each grid point is assigned a random value $\pm 1/\sqrt{\Delta V}$, where $\Delta V=V/N_{\text g}$ is the volume element and $V$ is the volume of the supercell.  Eq.~(\ref{eq:rho-sto}) can be rewritten in a compact form:
\begin{equation}
\rho(\mathbf{r})=2\langle|\xi(\mathbf{r})|^2\rangle_\chi,
\end{equation}
where $|\xi\rangle$, is a projected stochastic orbital:
\begin{equation}
|\xi\rangle=\sqrt{\hat{\rho}}|\chi\rangle=\sqrt{\theta_{\beta}(\hat{h}_{\text{KS}},\mu)}|\chi\rangle.
\label{eq:sto-orb}
\end{equation} 
The projection of the stochastic orbitals onto the occupied space is obtained by expanding $\sqrt{\theta_{\beta}(\hat{h}_{\text{KS}},\mu)}$ in Chebyshev polynomials:\cite{Kosloff1988,Kosloff1994} 
\begin{equation}
\sqrt{\theta_{\beta}(\hat{h}_{\text{KS}},\mu)}=\sum_{n=0}^{N_\text{c}} a_n(\mu,\beta)T_n(\hat{h}_{\text{KS}}),
\label{eq:cheby}
\end{equation}
where $N_c$ is the length of the Chebyshev polynomial expansion, $a_n(\mu,\beta)$ are the expansion coefficients, and $T_n$ are the Chebyshev polynomials of order $n$.

Ground state observables corresponding to any one-body operator, $\hat{O}$, can be evaluated using a similar stochastic trace formula: 
\begin{equation}
O=2\mathrm{Tr}(\hat{\rho}\hat{O})=2\langle\langle\xi|\hat{O}|\xi\rangle\rangle_\chi.
\end{equation}
Since the exact electron density can only be recovered by averaging infinitely many stochastic orbitals, estimates for $O$ result in a statistical error that decreases as $N_{\chi}^{-1/2}$ according to the central limit theorem, where, as before, $N_{\chi}$ is the number of stochastic orbitals. To achieve chemical accuracy for the electron density and the forces on the nuclei, $N_{\chi}$ may need to exceed $10^3$ orbitals, limiting the efficiency of sDFT. The need to develop noise reduction schemes is clear and would extend the range of system sizes that can be studied routinely using sDFT.

\section{Noise reduction schemes in stochastic DFT}
\label{sec:nrs}
\subsection{Overlapped Embedded Fragmented Stochastic DFT}
\label{sec:oefsdft}
Significant reduction in the statistical error can be achieved by introducing a reference system that provides a reasonable approximation to the electron density and can be calculated within KS-DFT. The full electron density is then given as a sum of the reference system electron density and a small correction term obtained stochastically.\cite{doi:10.1063/1.4890651,
doi:10.1063/1.4984931,doi:10.1002/wcms.1412,doi:10.1063/1.5064472}  
In this section we will briefly review the most recent developments based on an overlapped embedded fragmented stochastic DFT (o-efsDFT), which is central to the proposed ew-efsDFT. Full details of the approach can be found elsewhere.\cite{doi:10.1063/1.5064472}

\begin{figure}[t]
\centering \includegraphics[width=8cm]{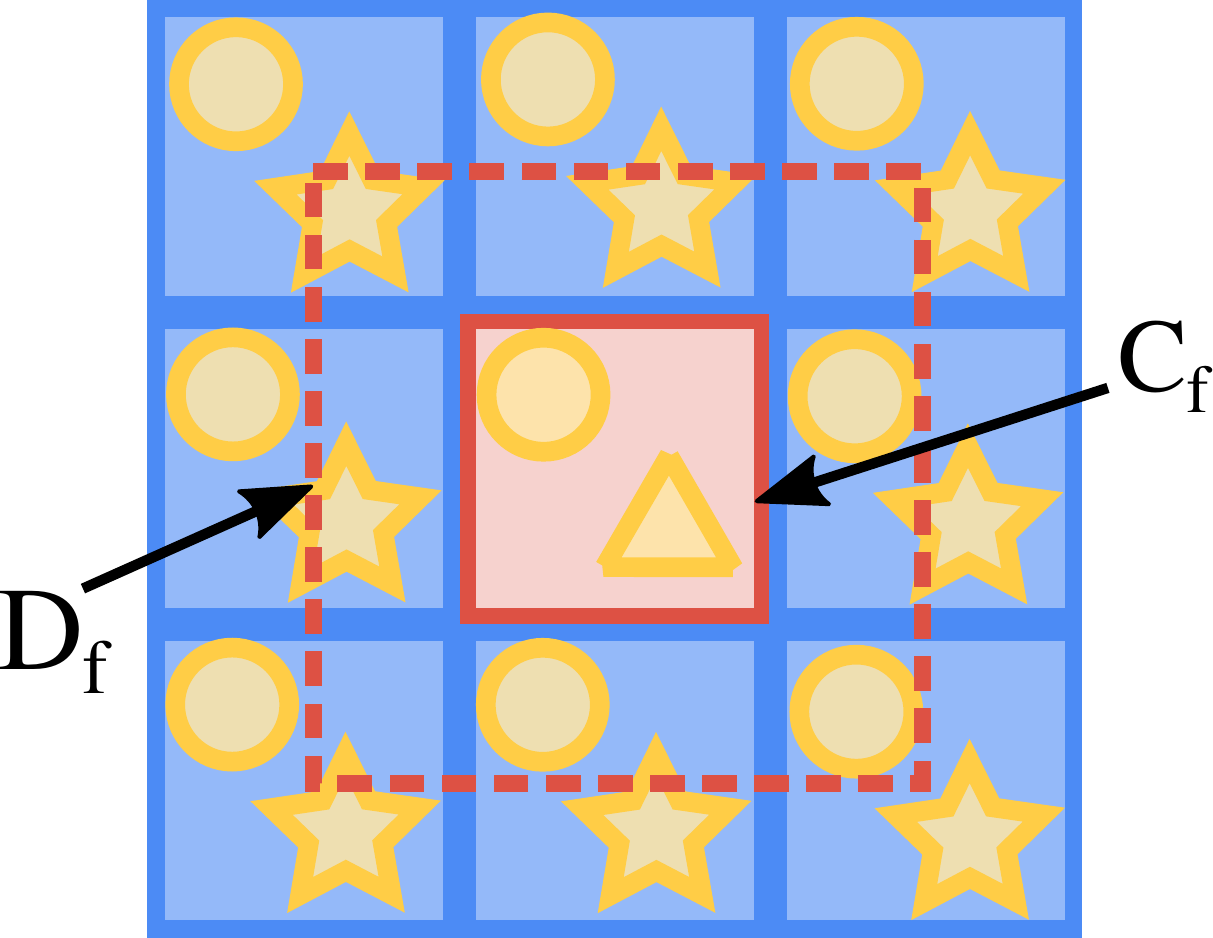}
\caption{An illustration of overlapped fragmented scheme. Each small solid blue/red square represents a core region $C_f$. We want to emphasis that the red square is different from blue squares by switching a star to a triangle, which is analogue to a impurity in solid. If we focus on the solid red square core region of a fragment, the region within the dashed red square corresponds to the dressed fragment, $D_f$. The region between the solid red square and dashed red square is the buffer region. }
\label{fig:bound}
\end{figure}

In o-efsDFT the supercell is divided into fragments referred to as "core regions'' (see Fig.~\ref{fig:bound} for an illustration) $C_f$ wrapped by "buffer regions'' ($B_f$) to form dressed fragments ($D_f=C_f\cup B_f$) where $f$ is the fragment index. The fragment density matrix, $\hat{\rho}_{f}$, is given by:
\begin{equation}
\langle\mathbf{r}|\hat{\rho}_{f}|\mathbf{r}'\rangle = 
\begin{cases}
\sum_{i=1}^{N_{\text{occ}}^f}\langle\mathbf{r}|\varphi^f_i\rangle\langle\varphi^f_i|
\mathbf{r}'\rangle & \;\;\mathbf{r}'\in D_f \\
0 & \;\;\mathbf{r}'\notin D_f
\end{cases}
\end{equation}
for $\mathbf{r}\in C_f$.  In the above equation, $\varphi^f(\mathbf{r})$ are the KS  orbitals for fragment $f$ obtained from a deterministic KS-DFT approach and $N_{\text{occ}}^f$ is the total number of occupied orbitals for the $f$'th dressed fragment. Using the above, the total electron density can be evaluated as follows:
\begin{align}
\rho(\mathbf{r}) & = 2\sum_f \langle\mathbf{r}|\hat{\rho}_{f}
\hat{\rho}_{f}^\top|\mathbf{r}\rangle+
2\langle|\xi(\mathbf{r})|^2\rangle_\chi \nonumber \\
& \quad -2\sum_f \left\langle\langle\mathbf{r}|\hat{\rho}_{f}|
\chi\rangle\langle\chi|\hat{\rho}_{f}^\top|\mathbf{r}\rangle
\right\rangle_\chi \nonumber \\
& = 2\sum_f \rho_f(\mathbf{r})+2\langle|\xi(\mathbf{r})|^2\rangle_\chi
-2\sum_f \langle|\xi_f(\mathbf{r})|^2\rangle_\chi,
\label{eq:rho-frag}
\end{align}
where the fragment electron density is $\rho_f(\mathbf{r})=\sum_{i=1}^{N_{\text{occ}}^f}|\varphi^f_i(\mathbf{r})|^2$, $\xi_f(\mathbf{r})=\sum_{i=1}^{N_{\text{occ}}^f}\varphi^f_i(\mathbf{r}) \langle\varphi^f_i|\chi\rangle_{D_f}$ and $\langle\varphi^f_i|\chi\rangle_{D_f}=\int_{D_f}\mathrm{d}\mathbf{r}{\varphi^f_i}(\mathbf{r})^\ast\chi(\mathbf{r})$. 
We use the relationship $\rho_f(\mathbf{r})=\langle\mathbf{r}|\hat{\rho}_{f}
\hat{\rho}_{f}^\top|\mathbf{r}\rangle$ in Eq.~(\ref{eq:rho-frag}) since KS-DFT methods 
in the 0K limit are adopted to calculate fragment KS orbitals. 

In the limit $N_\chi \rightarrow \infty$, the first and last terms on the right hand side of Eq.~(\ref{eq:rho-sto}) cancel, while the remaining term converges to the deterministic electron density. For a finite set of stochastic orbitals, the noise in the second term on the right hand side of Eq.~(\ref{eq:rho-sto}) roughly cancels that in the last term, as long as the reference system 
density matrix provides a reasonable approximation to that of the full system, thereby, leading to a significant reduction in the statistical error.
\cite{doi:10.1063/1.5064472,doi:10.1063/1.4890651}

\subsection{Energy Window Stochastic DFT}
\label{sec:ew-sdft}
A reduction in the statistical fluctuations can also be achieved using another scheme, based on partitioning the occupied space into ``energy windows''.\cite{doi:10.1063/1.5114984} In this approach, rather than projecting $|\chi\rangle$ onto the occupied space using Eq.~(\ref{eq:sto-orb}), we divide the occupied space into energy windows, and $|\chi\rangle$ is projected onto each window using a set of projectors, $\hat{\mathbf{P}}_1,\cdots,\hat{\mathbf{P}}_{N_{\text w}}$ (see Eq.~(\ref{eq:p})), that are calculated simultaneously with a single Chebyshev expansion.\cite{doi:10.1063/1.5114984} The electron density is given by the sum of all projected densities:
\begin{equation}
\rho(\mathbf{r}) = 2\sum_{w=1}^{N_{\text w}}\left\langle\left|{\xi^{(w)}} 
(\mathbf{r})\right|^2\right\rangle_\chi \equiv  2\sum_{w=1}^{N_{\text w}} \rho^{(w)}({\bf r}),
\end{equation}
where $|\xi^{(w)}\rangle=\sqrt{\hat{\mathbf{P}}_w}|\chi\rangle$ is a projected stochastic orbital for window $w$. The variance of the electron density in this scheme is $8\sum_{w=1}^{N_{\text w}}\left[\rho^{(w)}(\mathbf{r})\right]^2$, which is smaller than variance in sDFT given by $8\left(\sum_{w=1}^{N_{\text w}}\rho^{(w)}(\mathbf{r})\right)^2$.
\cite{doi:10.1063/1.5114984}

\section{Energy window embedded fragmented stochastic DFT}
\label{sec:ew-efsdft}
Combing the energy window approach with the fragmentation approach results in the following expression for the electron density at a grid point $\mathbf{r}$: 
\begin{widetext}
\begin{align}
\rho(\mathbf{r}) & =2\sum_f \rho_f(\mathbf{r})\nonumber \\
& \quad +\sum_{w=1}^{N_{\text w}}\left(2\left\langle\langle\mathbf{r}|
\sqrt{\hat{\rho}\hat{\mathbf{P}}_w}|\chi\rangle\langle\chi|\sqrt{\hat{\mathbf{P}}_w\hat{\rho}}|\mathbf{r}\rangle\right\rangle_\chi-2\sum_f \left\langle\langle\mathbf{r}|\hat{\rho}_{f} \sqrt{\hat{\mathbf{P}}_w}|\chi\rangle\langle\chi|\sqrt{\hat{\mathbf{P}}_w}\hat{\rho}_{f}^\top|\mathbf{r} \rangle\right\rangle_\chi\right) \nonumber \\
& = 2\sum_f \rho_f(\mathbf{r})+2\sum_{w=1}^{N_{\text w}}\left\langle|\zeta^{(w)}(\mathbf{r})|^2\right\rangle_\chi -2\sum_f \sum_{w=1}^{N_{\text w}}\left\langle|\xi^{(w)}_f(\mathbf{r})|^2\right\rangle_\chi.
\label{eq:rho-ew-frag}
\end{align}
\end{widetext}
In the above equation, the projection operators on the energy windows are defined as
\begin{equation}
\label{eq:p}
\begin{split}
&\hat{\mathbf{P}}_w = \theta_{\beta}(\hat{h}_{\text{KS}},\varepsilon_w)-\theta_{\beta}(\hat{h}_{\text{KS}},\varepsilon_{w-1}) \;\; 1\leq w<N_{\text w} \\
&\hat{\mathbf{P}}_{N_{\text w}} = \hat{I}-\sum_{w=1}^{N_{\text w}-1}\hat{\mathbf{P}}_w,
\end{split}
\end{equation}
where $\{\varepsilon\}\equiv \varepsilon_0 \cdots \varepsilon_{N_{{\rm w}-1}}$ 
($\varepsilon_0=-\infty$) define the boundaries of each energy windows and the action of $\sqrt{\hat{\rho}\hat{\mathbf{P}}_w}$ and $\sqrt{\hat{\mathbf{P}}_w}$ on $|\chi\rangle$ is obtained using a proper Chebyshev series:
\begin{equation}
\begin{split}
|\zeta^{(w)}\rangle &= \sqrt{\hat{\rho}\hat{\mathbf{P}}_w}|\chi\rangle=\\&\sum_{n=0}^{N_{\text c}}a_n^{(w)}(\mu,\varepsilon_w,\varepsilon_{w-1}) T_n(\hat{h}_{\text{KS}})|\chi\rangle \\
 |\xi^{(w)}\rangle &= \sqrt{\hat{\mathbf{P}}_w}|\chi\rangle=\\ & \sum_{n=0}^{N_{\text c}}b_n^{(w)}(\varepsilon_w,\varepsilon_{w-1}) T_n(\hat{h}_{\text{KS}})|\chi\rangle.
\end{split}
\end{equation}
Finally, as before, the density of each fragment is given by  $\rho_f(\mathbf{r})=\sum_{i=1}^{N_{\text{occ}}^f}|\varphi^f_i(\mathbf{r})|^2$ and the stochastic projected orbitals for each fragment are given by:
\begin{equation}
\xi_f^{(w)}(\mathbf{r}) =\sum_{i=1}^{N_{\text{occ}}^f}  \varphi^f_i(\mathbf{r}) \langle \varphi^f_i|\xi^{(w)}\rangle_{D_f}.
\end{equation} 

For all applications below, 
% Ming change we divide the energy space into $N_{\rm w}=41$ windows, where 
$\varepsilon_w$ are held fixed for the entire self-consistent procedure and are independent of the chemical potential, $\mu$. Other choices  of the window boundaries can affect the level of noise;  however, significant simplicity is achieved for fixed window boundaries.  In this case, the chemical potential can be obtained by solving:
\begin{align}
N(\mu)= & 2 \sum_f\int_{C_f}\mathrm{d}\mathbf{r}\rho_f(\mathbf{r})+2\left\langle\langle\chi|\hat{\rho}(\mu)|
\chi\rangle\right\rangle_\chi \nonumber \\
& -2\sum_f\sum_{w=1}^{N_{\text w}}\left\langle\int_{C_f}\mathrm{d}\mathbf{r}|\xi^{(w)}_f(\mathbf{r})|^2
\right\rangle_\chi,
\label{eq:num-elec}
\end{align}
where $\int_{C_f}\mathrm{d}\mathbf{r}$ imply that the integrals are preformed in real-space in region ${\bf r} \in C_f$.  In the above equation, $\langle\chi|\hat{\rho}(\mu)|\chi\rangle$ is evaluated by expanding $\hat{\rho}$ in a Chebyshev series $\langle\chi|\hat{\rho}(\mu)|\chi\rangle=\sum_{n=0}^{N_{\text{c}}}c_n(\mu)\langle\chi|T_n(\hat{h}_{\text{KS}})|\chi\rangle$ and the chemical potential is determined by solving for $N(\mu^{*})=N_{\text e}$ where $N_{\text e}$ is the total number of electrons in the system. 

Similar to the electron density, other ground state observables such as the kinetic energy 
\begin{align}
E_{\text k} & = \nonumber \\
& 2\sum_f\sum_{i=1}^{N_{\text{occ}}^f}\langle\varphi^f_i|\hat{t}|
\varphi^f_i\rangle_{C_f}+
2\sum_{w=1}^{N_{\text w}}\left\langle\langle\zeta^{(w)}|\hat{t}|\zeta^{(w)}\rangle\right\rangle_\chi \nonumber \\
& \quad -2\sum_{i=1}^{N_{\text w}}\sum_f\left\langle\langle\xi_f^{(w)}|\hat{t}|
\xi_f^{(w)}\rangle_{C_f}\right\rangle_\chi,
\label{eq:ek-ew-frag}
\end{align}
or the non-local pseudopotential energy
\begin{align}
E_{\text {nl}} & =  2\sum_f\sum_{I,\mathbf{R}_I\in C_f}\sum_{i=1}^{N_{\text{occ}}^f} 
\langle\varphi^f_i|\hat{v}_{\text{nl}}^{I}|\varphi^f_i\rangle_{D_f} \nonumber \\
& \quad +2\sum_{w=1}^{N_{\text w}}\sum_I
\left\langle\langle\zeta^{(w)}|\hat{v}_{\text{nl}}^{I}|\zeta^{(w)}
\rangle\right\rangle_\chi \nonumber \\
& \quad -2\sum_{w=1}^{N_{\text w}}\sum_f\sum_{I,\mathbf{R}_I\in C_f}
\left\langle\langle\xi_f^{(w)}|\hat{v}_{\text{nl}}^{I}|
\xi_f^{(w)}\rangle_{C_f}\right\rangle_\chi,
\label{eq:enl-ew-frag}
\end{align}
or the non-local pseudopotential contribution to the forces on the nuclei
\begin{align}
\mathbf{F}_{\text {nl}}^I = & 2\sum_{i=1}^{N_{\text{occ}}^f}\left
\langle\varphi^f_i\middle|\frac{\partial\hat{v}_{\text{nl}}^{I}}{\partial \mathbf{R}_I}
\middle|\varphi^f_i\right\rangle_{D_f} \nonumber \\
& +2\sum_{w=1}^{N_{\text w}}\left\langle\left\langle\zeta^{(w)}\middle|
\frac{\partial\hat{v}_{\text{nl}}^{I}}{\partial \mathbf{R}_I}
\middle|\zeta^{(w)}\right\rangle\right\rangle_\chi \nonumber \\
& -2\sum_{w=1}^{N_{\text w}}\left\langle\left\langle\xi_f^{(w)}\middle|
\frac{\partial\hat{v}_{\text{nl}}^{I}}{\partial \mathbf{R}_I}
\middle|\xi_f^{(w)}\right\rangle_{D_f}\right\rangle_\chi,
\label{eq:fnl-ew-frag}
\end{align}
is expressed in ew-ofsDFT as a sum of three terms, where the first term and the last term cancel each other in the limit $N_{\chi }\rightarrow \infty$ (see supplementary material). 

The proposed ew-efsDFT method to reduce the noise in the density, energy, and forces on the nuclei can be summarized as follows:
\begin{enumerate}
\item Generate the KS orbitals $\{\varphi_i^f(\mathbf{r})\}$ for each dressed fragment and $\rho_f(\mathbf{r})=\sum_{i=1}^{N_{\rm occ}^f} |\varphi_i^f({\bf r})|^2$ using a deterministic DFT. $\rho({\bf r}) = 2\sum_f\rho_f(\mathbf{r})$ is used as the initial electron density guess.

\item For each stochastic orbital $\chi({\bf r})$ (defined above),  calculate and store on the grid the projected stochastic orbital $\zeta^{(w)}({\bf r})=\langle {\bf r} | \sqrt{\hat{\mathbf{P}}_w|}\chi\rangle$ and also store the Chebyshev moments, $\langle\chi|T_n(\hat{h}_{\text{KS}})|\chi\rangle$. 

\item For each window and for each stochastic orbital, generate and store on the grid $\xi_f^{(w)}(\mathbf{r}) =\sum_{i=1}^{N_{\text{occ}}^f}\varphi^f_i(\mathbf{r})\langle\varphi^f_i|\xi^{(w)}\rangle_{D_f}$. 

\item Solve for $\mu^*$  ($N(\mu^*)=N_{\text e}$)  with the {\it regula falsi} method using the Chebyshev moments ($\langle\chi|T_n(\hat{h}_{\text{KS}})|\chi\rangle$), $\xi_f^{(w)}(\mathbf{r})$, and $\rho_f(\mathbf{r})$.\cite{doi:10.1002/wcms.1412}

\item For each window and for each stochastic orbital, generate and store on the grid the stochastic projected orbitals $\zeta^{(w)}({\bf r})=\langle {\bf r} | \sqrt{\hat{\rho}(\mu^*)\hat{\mathbf{P}}_w} | \chi \rangle$ using the chemical potential determined in the previous step. 

\item Generate and store the electron density $\rho(\mathbf{r})$ using Eq.~(\ref{eq:rho-ew-frag}) with all stochastic orbitals.

\item Update the density and the KS Hamiltonian using the iterative subspace (DIIS) method~\cite{PULAY1980393} and repeat the above steps until self-consistency is achieved, using the same random number seed. 
\end{enumerate}

\section{Application to G-Center in Bulk Silicon}
\label{sec:results}
\begin{figure}[t]
\centering \includegraphics[width=7cm]{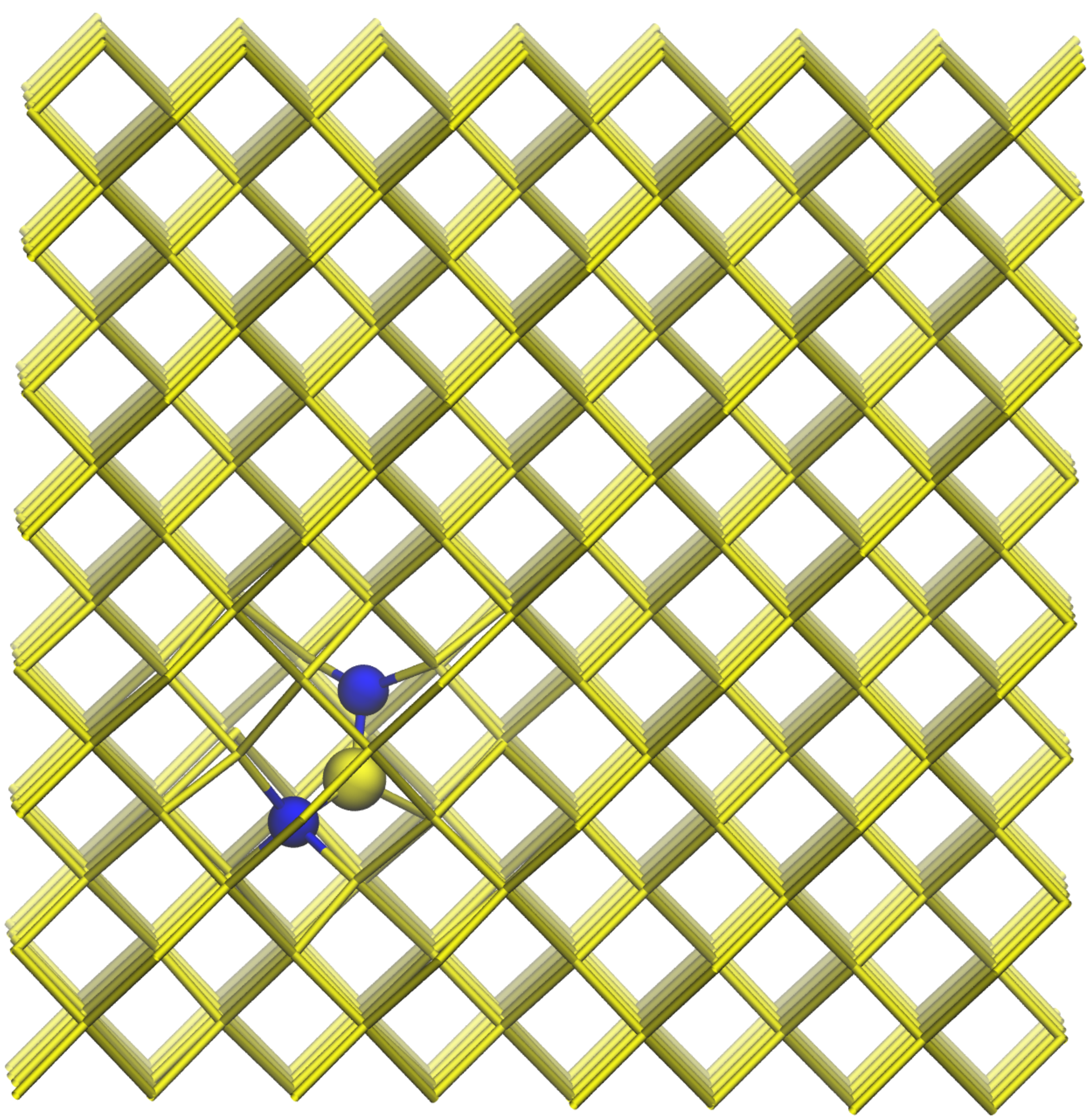}
\caption{An A-type G-center (two carbon atoms shown in blue) embedded in a Si$_{512}$ supercell (Si-Si bonds shown in yellow). The A-type G-center is constituted by a substitutional carbon atom, an interstitial carbon atom, and an interstitial silicon atom, which are highlighted as spheres.}
\label{fig:gc-structure}
\end{figure}

\begin{figure}[t]
\centering \includegraphics[width=7cm]{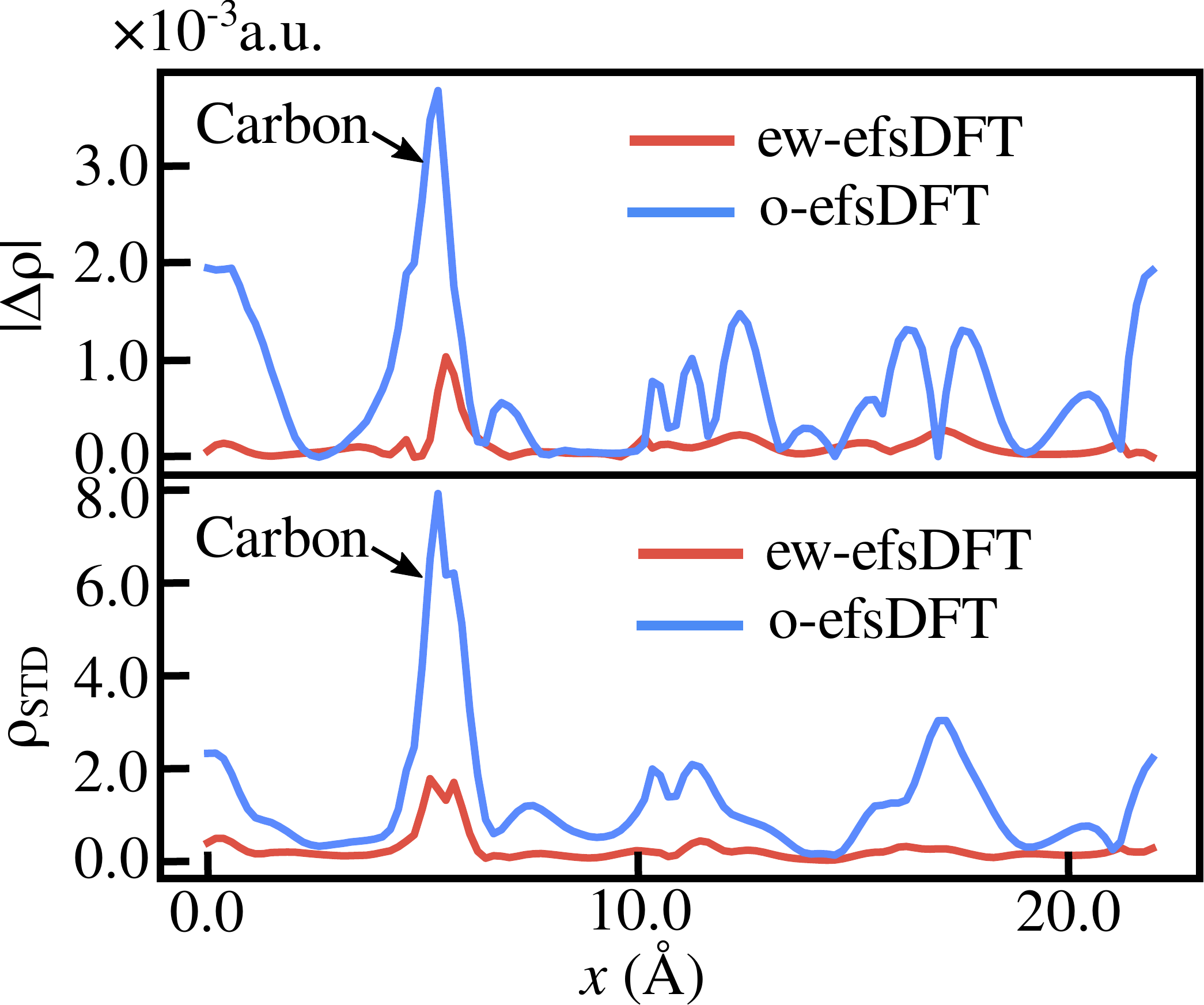}
\caption{Upper panel: The absolute value of the electron density difference ($|\Delta\rho|$) between the stochastic (ew-efsDFT in red, o-efsDFT in blue) and a deterministic calculation. Lower panel: The standard deviation of electron density ($\rho_{\text{STD}}$) evaluated with ew-efsDFT (red line) and o-efsDFT (blue line). $|\Delta\rho|$ and $\rho_{\text{STD}}$ are shown along the $x$ axis with $y=5.3$\AA and $z=5.0$\AA. The peak marked by ``Carbon'' corresponds to a region closed to a carbon atom.}
\label{fig:rho}
\end{figure}

\begin{table*}[t]
\begin{tabular}{cllllll}
\hline
Method & $E_{\text{k}}/N_{e}$  & $E_{\text{nl}}/N_{e}$  & $E_{\text{H}}/N_{e}$  & $E_{\text{loc}}$ & $E_{\text{xc}}$ & $E_{\text{tot}}/N_{\text{e}}$  \tabularnewline
\hline
dDFT  & 10.1767 & 5.7871 & 2.0194  & -8.7085 & -8.1027 & -26.8197 \tabularnewline
o-efsDFT  & 10.1735(35) & 5.7882(36) & 2.0222(19)  & -8.7074(40) & -8.1052(8) & -26.8205(11) \tabularnewline
ew-efsDFT  & 10.1778(11) & 5.7874(4) & 2.0193(6)  & -8.7086(11) & -8.1028(3) & -26.8186(4) \tabularnewline
\hline
\end{tabular}
\caption{\label{table:energy} The kinetic energy ($E_{\text{k}}$), the nonlocal pseudopotential energy ($E_{\rm nl}$), the Hartree energy ($E_{\rm H}$), the local pseudopotential energy ($E_{\rm loc}$), the exchange-correlation energy ($E_{\rm xc}$), and the total energy ($E_{\rm tot}$), all per electron in eV obtained by a deterministic DFT (dDFT), o-efsDFT, and ew-efsDFT. The standard deviation is presented in parenthesis.}
\end{table*}

We demonstrate the ew-efsDFT for low G-center defect concentration in bulk silicon.\cite{PhysRevB.42.5765,doi:10.1063/1.2766843}  We focus on the A-type~\cite{PhysRevB.42.5765} G-center impurity embedded within a Si$_{512}$ supercell (see the structure in  Fig.~\ref{fig:gc-structure}).  We performed $\Gamma$ point DFT calculations using the Perdew-Burke-Ernzerhof (PBE)~\cite{PhysRevLett.77.3865} functional with Troullier-Martins norm-conserving pseudopotentials~\cite{PhysRevB.43.1993} in the Kleinman-Bylander form.\cite{PhysRevLett.48.1425} As a result of localized orbitals around the carbon atoms, a $40$Ryd wave function cutoff ($80$Ryd for the density cutoff) was used, corresponding to real space grid spacing of $0.18$\AA. In-gap states require a large $\beta\approx 900$ in inverse Hartree to sufficiently converge the ground state properties with respect to the electron temperature. $80$ stochastic orbitals were used in both ew-efsDFT and o-efsDFT. The dressed fragments were Si$_{64}$ with periodic boundary conditions while the size of each core region was Si$_8$.  As pointed out above, $41$ energy windows were used in ew-efsDFT with a width that is inversely proportional to the density of states at the center of the window. This guarantees that each energy windows contains roughly the same number of KS orbitals, thereby lowering the statistical noise.

\subsection{Results}
In  Fig. \ref{fig:rho} we assess the accuracy of ew-efsDFT for the electron density (upper panel) and the standard deviation in the electron density (lower panel) for selected positions in the vicinity of the G-center. The density and standard deviation were calculated from $5$ independent ew-efsDFT or o-efsDFT runs, with $80$ stochastic orbitals for each run. For clarity, we plot the absolute value of the electron density difference between the stochastic and the deterministic calculations. Both the deviations from the deterministic electron density and the standard deviation obtained by the ew-efsDFT (shown in red) are significantly smaller than the corresponding o-efsDFT results (shown in blue). The noise of the electron density is  significantly smaller,  by approximately a factor of $5$ when compared to o-efsDFT.

In Table~\ref{table:energy} we list the kinetic energy ($E_{\text{k}}$), the nonlocal pseudopotential energy ($E_{\rm nl}$), the Hartree energy ($E_{\rm H}$), the local pseudopotential energy ($E_{\rm loc}$), the exchange-correlation energy ($E_{\rm xc}$), and the total energy ($E_{\rm tot}$), all per electron. The reference deterministic calculation is converge (on the grid) to all significant digits shown. In parenthesis we provide the standard error, which is significantly smaller in ew-efsDFT compared to o-efsDFT for all quantities. We find that the standard error in the total energy per electron decreased by a factor of $\approx 3$ when 41 windows were used.  The total energy per electron in both ew-efsDFT and o-efsDFT are slight outside one standard deviation from the deterministic DFT result.\cite{PhysRevB.97.115207,doi:10.1002/wcms.1412} We note in passing that the ew-sDFT approach (without fragmentation) does not reduce the noise in the total energy per electron, as discussed previously.\cite{doi:10.1063/1.5114984} 

\begin{figure}[h]
\centering \includegraphics[width=7cm]{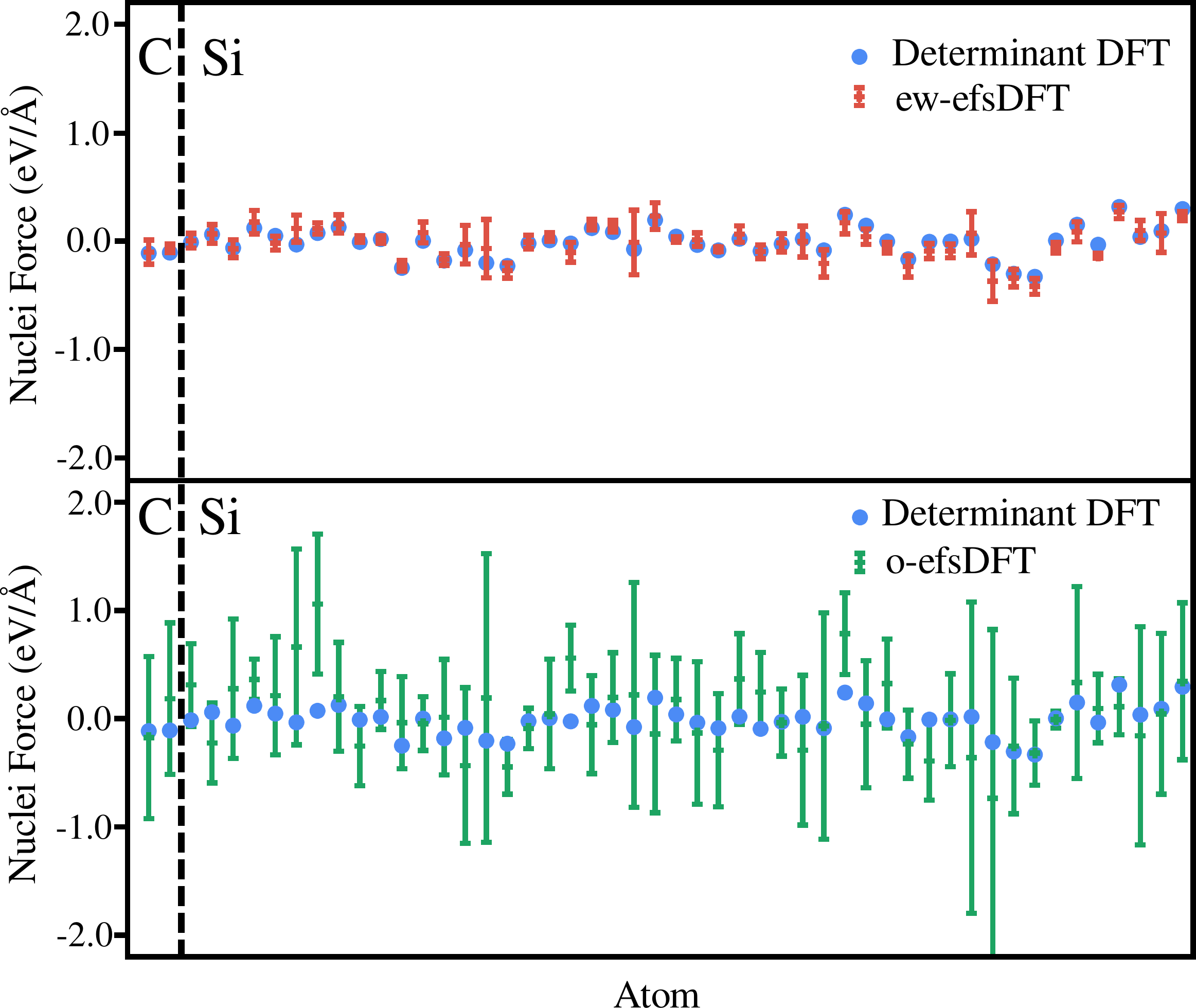}
\caption{Forces on the nuclei along the $x$-direction (${F}_x$) for selected atoms calculated from ew-efsDFT (upper panel) and o-efsDFT (lower panel). Error bars in the forces on the nuclei were obtained from $5$ runs. Blue symbols are ${F}_x$ calculated from a deterministic DFT.  The dashed line signifies the boundary between carbon and silicon atoms}
\label{fig:force}
\end{figure}

In Fig.~(\ref{fig:force}) we plots the force on the nuclei along the $x$-direction for selected atoms obtained by the ew-efsDFT (upper panel) and the o-efsDFT (lower panel). Error bars indicate the standard deviation for each force. Clearly the statistical fluctuations are significantly smaller for ew-efsDFT compared to o-efsDFT. In order to estimate the overall noise reduction efficiency, we averaged the standard deviations of ${F}_x$ (nuclei force along $x$ axis) over all atoms. The averaged standard deviation of ${F}_x$ is $\approx 0.53$eV/\AA~~for o-efsDFT and $\approx 0.09$eV/\AA~~for ew-efsDFT. Similar results were also obtained for other force components ${F}_y$ and ${F}_z$, which implies that  ew-efsDFT standard deviations of nuclei forces is about a factor of $6$ smaller than that in o-efsDFT. In other words, to achieve similar noise level in o-efsDFT would require $\approx 30$ times more stochastic orbitals. No bias was observed for forces on the nuclei in ew-efsDFT.

\subsection{Computational Cost}
\begin{table}[t]
\begin{tabular}{ccc}
\hline
& o-efsDFT & ew-efsDFT \tabularnewline
\hline
Wall time (h) & 25.11 & 25.22 \tabularnewline
\hline
Number of \\ SCF Iterations & 25.6 & 16.4 \tabularnewline
\hline
Time for One \\ SCF Iteration (h) & 0.98 & 1.54 \tabularnewline
\hline
Time for Calculating \\ $\langle\chi|T_n(\hat{h}_{\text{KS}})|\chi\rangle$ (h)
& 0.32 & 0.73 \tabularnewline
\hline
Time for Generating \\ $|\xi\rangle$ or $|\xi^{(w)}\rangle$ (h) 
& 0.64 & 0.72 \tabularnewline
\hline
Time for projecting \\ $|\zeta^{(w)}\rangle$ with $\hat{\rho}_{\text{frag}}$ (h) 
& N/A & 0.04 \tabularnewline
\hline
\end{tabular}
\caption{Averaged computational time (in hours) for o-efsDFT and ew-efsDFT. The wall-time and number of SCF iterations are averaged over $5$ independent runs. Time of calculating $|\xi^{(w)}\rangle$, $|\zeta^{(w)}\rangle$, and $|\xi\rangle$ and time of projecting $|\xi^{(w)}\rangle$ with $\hat{\rho}_{\text{frag}}$ are averaged over all SCF iterations and all $5$ runs. All calculations were performed on a $40$ node cluster computer, where each node contains two $16$-core Intel Xeon Processors E5-2698 v3 at 2.3GHz.}
\label{table:time}
\end{table}

In Table~\ref{table:time} we summarize the total wall time and the contribution from the Chebyshev moments, $\langle\chi|T_n(\hat{h}_{\text{KS}})|\chi\rangle$, and from the projections of $|\chi\rangle$ onto the occupied space and the energy windows. The number of stochastic orbitals, the size of the core and dressed fragments, the grid size, and all other parameters were the same for both approaches.  The overall wall time seems to be very similar, comparing ew-efsDFT and o-efsDFT methods.  Each SCF iteration is $\approx 50\%$ longer in ew-efsDFT, but the number of SCF iterations required to achieve a similar convergence is smaller in ew-efsDFT, resulting in similar wall  times.  Note that the statistical error in ew-efsDFT is much smaller than o-efsDFT. To achieve similar statistical errors in o-efsDFT would result in wall times that are roughly $30$ times longer than ew-efsDFT.

The main difference between the two methods is the computational time for generating the Chebyshev moments, $\langle\chi|T_n(\hat{h}_{\text{KS}})|\chi\rangle$. In o-efsDFT we used the relation $T_{2n}=2T_n^2-1$ to evaluate $T_n(\hat{h}_{\text{KS}})|\chi\rangle$ for $n>N_c/2$, thereby computing only $1/2$ the number of moments.  This relation cannot be used in ew-efsDFT since each stochastic orbital is projected onto all energy window to generate $| \xi^{(w)}\rangle$ and $|\zeta^{(w)}\rangle$.  Other smaller contributions to the computational time difference between the two methods can be traced to the need to generate $N_w$ projected stochastic orbitals in ew-efsDFT compared to only one projected orbital in o-efsDFT, resulting in $\approx 20\%$ increase in generating $|\xi\rangle$ vs. $|\xi^{(w)}\rangle$.  In addition, in ew-efsDFT one has to compute $|\xi_f^{(w)}\rangle$ with the fragment density matrix, i.e. $\sum_{i=1}^{N_{\text{occ}}^f}\varphi^f_i(\mathbf{r})\langle\varphi^f_i| \xi^{(w)}\rangle_{D_f}$,  in each SCF iteration while in o-efsDFT projection of $|\chi\rangle$ is preformed only once at the beginning of the calculation.

\section{Summary}
\label{sec:conclusions}
In this work, we have developed an approach to reduce the statistical fluctuations in the electron density, total energy, and in the forces on the nuclei within the stochastic DFT framework,  without increasing the number of stochastic orbitals.  This achievement was made possible by combining the overlapped embedded-fragmented stochastic DFT~\cite{doi:10.1063/1.5064472} with the energy windows stochastic DFT.\cite{doi:10.1063/1.5114984} The new approach builds on both real-space and energy-space fragmentation, resulting in a significant reduction of the noise in single-particle observables without affecting the computational time. The performance of the ew-efsDFT was tested for a G-center embedded in bulk silicon with a small fundamental gap and in-gap impurity states, making this a rather challenging system for DFT.  Compared to o-efsDFT and ew-DFT (not shown explicitly here),  the statistical error in the forces is approximately $6$ times smaller in ew-efsDFT resulting in a reduction of $\approx 30$ in the computational wall time. This reduction in noise/computational time is important to accurately describe structural properties of extended systems without the need to increase the number of stochastic orbitals. Application of the ew-efsDFT method to structural minimization are currently underway.\cite{structmin}

\section*{DATA AVAILABILITY}
The data that support the findings of this study are available from the corresponding author upon reasonable request.

%\section*{Supplementary Material}
%A discussion of the convergence on the electron density, energy, and forces on the nuclei as well as a discussion of the variance of the %electron density for ew-efsDFT are provided in the supplementary material. 

\acknowledgments
We acknowledge support from the Center for Computational Study of Excited State Phenomena in Energy Materials (C2SEPEM) at the Lawrence Berkeley National Laboratory, which is funded by the U.S. Department of Energy, Office of Science, Basic Energy Sciences, Materials Sciences and Engineering Division under Contract No. DE-AC02-05CH11231 as part of the Computational Materials Sciences Program. Computational resources were provided by the National Energy Research Scientific Computing Center (NERSC), a U.S. Department of Energy Office of Science User Facility operated under Contract No. DE-AC02-05CH11231. R.B. gratefully thanks the support of the Germany-Israel Foundation GIF grant number I-26-303.2-2018.

\bibliographystyle{aipnum4-1}
\bibliography{ew-frag}

\end{document}